\pdfoutput=1
\documentclass[11pt,onecolumn,letterpaper]{article}
\usepackage[T1]{fontenc}
\usepackage{times}
\usepackage{graphicx}
\usepackage{amsmath}
\usepackage{amssymb}
\usepackage[bf]{caption}
\usepackage{bm}
\usepackage{mathtools}
\usepackage[superscript,biblabel]{cite}
\usepackage{url}
\usepackage{booktabs}
\usepackage{multirow}
\usepackage[margin=2.5cm]{geometry}
\usepackage{xcolor}
\usepackage{setspace}


\let\oldcite\cite
\renewcommand{\cite}[1]{\unskip\oldcite{#1}}

\title{In Situ Training of Optical Neural Networks by Coding Agents}

\author
{Ilker Oguz$^{1,*}$, Colin Christian$^{2}$, Christophe Moser$^{2}$,\\ Demetri Psaltis$^{2}$ and Gordon Wetzstein$^{1,*}$\\
\\
\normalsize{$^{1}$Department of Electrical Engineering, Stanford University, Stanford, CA, USA}\\
\normalsize{$^{2}$School of Engineering, Institute of Electrical and Micro Engineering,\ }\\
\normalsize{Ecole polytechnique f\'ed\'erale de Lausanne, Lausanne, Switzerland}
}

\date{}

\begin{document}

\pagestyle{plain}
\thispagestyle{plain}
\maketitle
\vspace{-38.05pt}
\begin{center}
$^{*}$To whom correspondence should be addressed;\\
E-mail: ilkero@stanford.edu, gordon.wetzstein@stanford.edu
\end{center}
\vspace{0.2pt}
\setstretch{1.15}

\begin{quote}
{\bf
Optical neural networks promise fast, energy-efficient computing, but every device departs from its design and must be trained with the hardware. In situ training relies on an expert-built calibrated digital model of each device or on gradient-free trial and error. Coding agents, large language models that write and run programs, have shown an emergent ability to understand and edit the weights of digital neural networks. Here we introduce optics as code, a method in which a coding agent trains an optical neural network from measurements alone, through a harness that returns camera images of its probes and requires a hypothesis per request. We discovered that the agent forms a physical understanding of the optics and reparametrizes the network with interpretable optical elements tuned on the hardware. It outperforms gradient-free methods in classification and image encoding, approaching backpropagation through a calibrated model at a fraction of the cost. Needing neither per-device calibration nor gradients through the hardware, and improving with more capable models, optics as code offers a practical route to large, scalable optical neural networks and a general tool for calibrating and optimizing optical experiments.
}
\end{quote}

\begin{figure}[p]
    \centerline{\includegraphics[width=18cm]{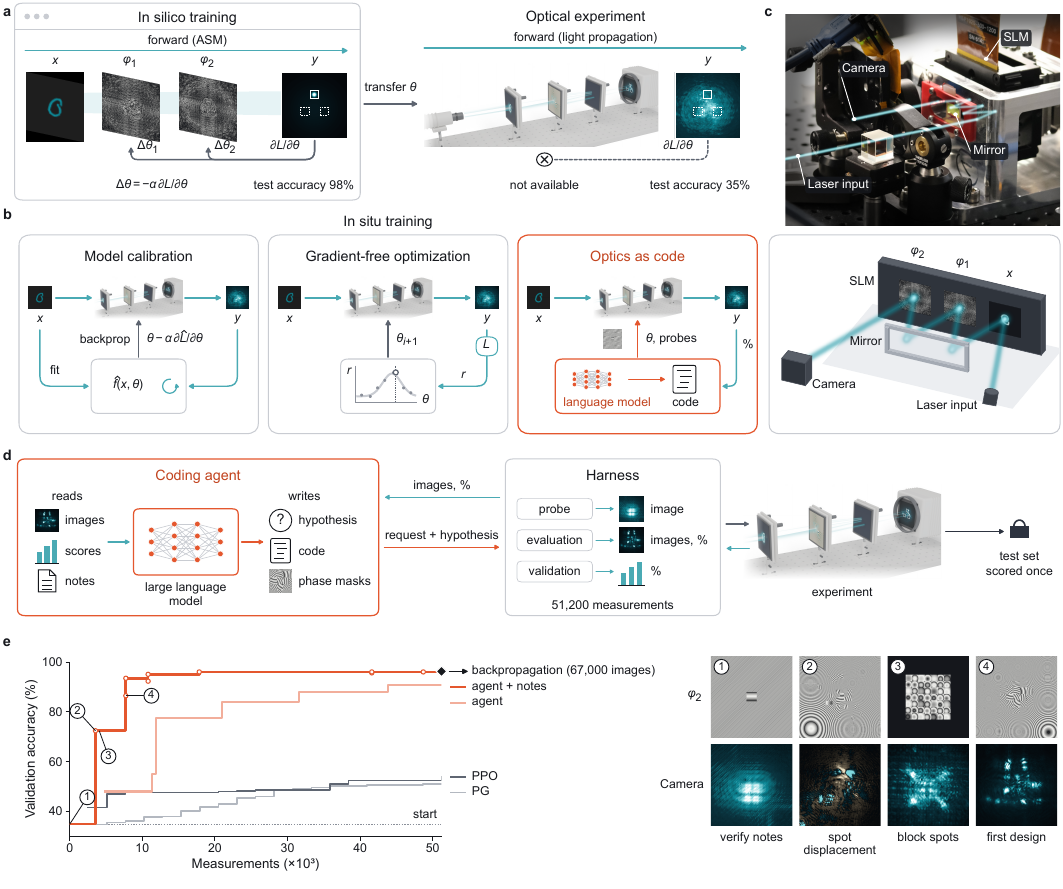}}
    \caption{\textbf{A coding agent trains an optical neural network in situ, from measurements alone.}
    \textbf{a}, Conventional deployment of optical neural networks. Phase masks trained in silico on an ideal model (angular-spectrum method, ASM) lose accuracy on the optical experiment, where gradients are not directly accessible.
    \textbf{b}, The two earlier routes to in situ training, and optics as code, in which a large language model writes code whose output, the phase masks $\theta$ and the probes, is displayed on the experiment. $\hat f$, calibrated model. $L$, loss. $r$, reward.
    \textbf{c}, Photograph and schematic of the experiment. SLM, spatial light modulator.
    \textbf{d}, The coding agent, a large language model, reads camera images, scores and its notes, and writes hypotheses, code, phase masks and notes of its findings. It reaches the experiment only through the harness, the software that executes its requests and returns the measurements. The test set is sealed.
    \textbf{e}, Best validation accuracy versus measurements on the experiment for the agent with the notes of an earlier session (circles, validations) and without, and for proximal policy optimization (PPO) and policy gradient (PG). Diamond, backpropagation through a calibrated model, whose calibration alone used 67,000 camera images (arrow, beyond the axis). Dotted, start. Right, $\varphi_2$ (top) and camera data (bottom) at numbered reasoning steps. 1, probes verifying the earlier notes. 2, the class displaces the focused spots (difference of two classes). 3, one spot per block of phase mask $\varphi_2$. 4, the agent's first own design from these spots.}
    \label{fig:first}
\end{figure}

Optical neural networks perform the operations of artificial intelligence (AI) with light, and they promise high speed and low energy consumption at large scale, because light propagates with low loss and carries many signals in parallel~\cite{r1,r2,r3,r5}. Their parameters are physical quantities, for instance the thickness of microfabricated or printed diffractive layers~\cite{r2} or the phase masks displayed on a spatial light modulator (SLM). Every assembled device deviates from the ideal functional form of the model it was designed with, through fabrication errors, the response of its modulators, misalignments between optical planes and aberrations~\cite{r8}. As a result, networks trained in simulation lose accuracy on the hardware, and they recover it only after fine-tuning on the device itself~\cite{r11,r12,r18,r21} (Fig.~\ref{fig:first}a).

To train a network on the hardware, where the gradients of the physical device are not directly accessible, two main routes have been developed, namely model calibration and gradient-free optimization~\cite{r9,r11,r12,r14,oguzff,r16,r17,r18,r19} (Fig.~\ref{fig:first}b). In model calibration, a differentiable model of the device, such as a calibrated angular-spectrum propagation model~\cite{r35,r36,yildirim2024} or a digital twin~\cite{r11,r12,oguz2025twin}, is first fitted to measured inputs and outputs, and the parameters are then trained by backpropagation through this model. Although this route reaches high accuracy, building and maintaining such a model for every device requires an expert, many measurements and substantial computation. Gradient-free optimization instead learns from the measured outputs alone, with methods that range from perturbative gradient estimates~\cite{r39} and evolutionary or Bayesian search~\cite{r33,r34} to forward-only training~\cite{oguzff,r16,r17} and reinforcement learning~\cite{r19,r41}, among which policy-based methods such as proximal policy optimization (PPO)~\cite{r20} have performed best on optical tasks~\cite{r21}. These methods, however, rely on trial and error, guided by generic statistical models fitted to the rewards alone, and the number of experiments they require rises steeply with the number of parameters. They become efficient only when a compact description of the optics is known, as in wavefront shaping~\cite{r29} and sensorless adaptive optics~\cite{r31}, and for an optical neural network with unknown errors this description must first be found by an expert. Hence, the amount of data and expert effort that in situ training requires remains one of the main challenges for deploying optical neural networks at scale.

This challenge calls for an optimizer that, like an experimental physicist, understands the system it tunes and chooses informative measurements. Large language models, the neural networks that underlie current AI assistants, are trained on text that includes scientific papers, textbooks, device manuals and code, and they thus combine knowledge of optics, instrumentation and programming otherwise spread over many experts. Used as coding agents, they act on this knowledge with the efficiency of a computer program, writing, running and revising their own programs over many steps without human intervention, and in this way they have discovered new algorithms~\cite{r22,r23,r24}, contributed to open problems in mathematics and physics~\cite{r43,r45} and operated laboratory instruments~\cite{r26,r27}, including the alignment of optical setups~\cite{lopez2026}. A recently reported ability of these models, weight-space programming, is of particular interest for neural networks~\cite{r25}. In this setting, an agent reads the weights of a separate digital network, forms an understanding of how the network works and edits or writes its weights, although it was never trained to do so. An independent study confirmed that these models can read small networks and write working ones as weights~\cite{goworek2026}. Such a model can thus modify a network directly, without the need for many small trial-and-error updates. The parameters of a physical network, however, can only be measured, and how an agent could understand and program such a device has remained an open question.

\enlargethispage{\baselineskip}Here we introduce optics as code, a method in which coding agents understand and program physical optical systems efficiently, from measurements alone. The agent works through a harness that we designed, the software interface that defines what it may request from the device and what it receives. We discovered that this framework enables in situ training of optical neural networks with high accuracy, and that the agent proceeds intelligently, in that it studies the optics, understands how they depart from their design and reparametrizes the network with a few physically interpretable parameters, which it then tunes on the hardware. The name reflects that the agent writes the optical function as a program, with lenses, prisms and aberration terms as its variables, and edits and tests this program on the device (Fig.~\ref{fig:first}b, right). We describe the experiment and the harness, benchmark the agent on two tasks, analyse its physical designs and identify which elements of the harness matter. Building on the notes of an earlier session, the agent trained an optical classifier to 96.8\% test accuracy, against at most 52.5\% for the gradient-free methods, and even exceeded backpropagation through a calibrated model of the device (96.3\%), at a fraction of its cost in data and training time. From 48 probe measurements, the agents also determined the magnifications, astigmatism and tilt sensitivities of the optical planes, in agreement with a calibrated model whose training required tens of thousands of calibration images. Since this ability requires no expert for each device and grows with the capability of the language models it builds on, optics as code offers a practical way to deploy optical neural networks in large numbers.

\begin{figure}[!t]
    \centerline{\includegraphics[width=18cm]{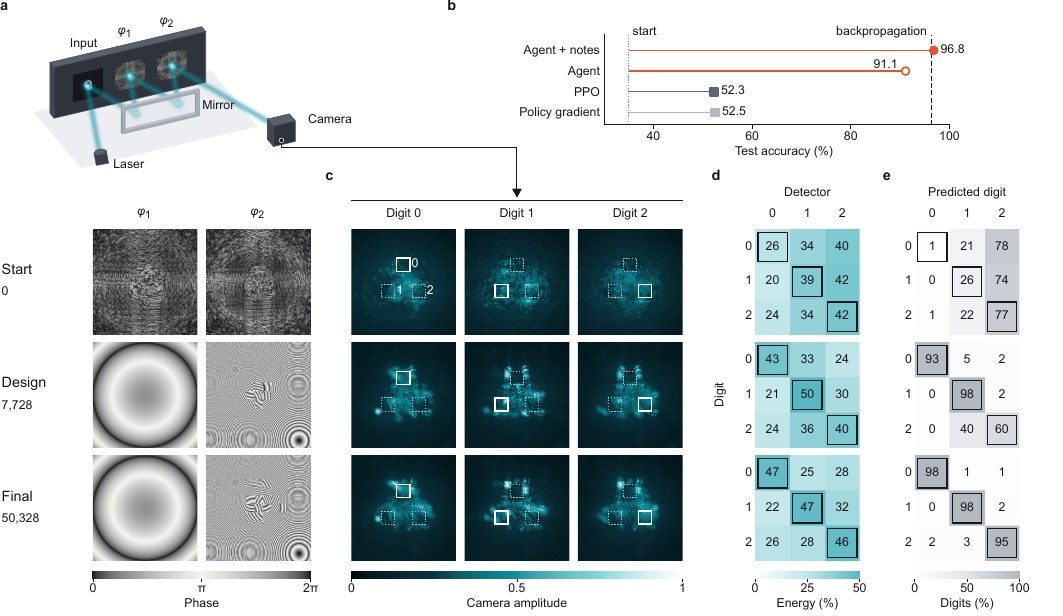}}
    \caption{\textbf{The agent trains an optical classifier on the experiment to the accuracy of backpropagation.}
    \textbf{a}, Schematic of the experiment, and the two phase masks at the start, at the agent's first own design and at the end of the best agent run, given the notes of an earlier session, with the measurements spent.
    \textbf{b}, Test accuracy on the 3,147 test digits at the end of the budget, for the best run of each method. Filled circle, agent with the notes of an earlier session. Open circle, agent without notes. Squares, PPO and policy gradient. Dashed, phase masks trained by backpropagation through a calibrated model of the device and scored on the experiment (96.3\%). Dotted, start.
    \textbf{c}, Camera images averaged over the practice digits of each class, at the three stages, with the three detectors.
    \textbf{d}, Fraction of the detected energy on each detector, per digit class.
    \textbf{e}, Confusion matrix, in per cent of the digits of each class. Start and final phase masks in \textbf{d} and \textbf{e}, 3,147 test digits. First design, 128 practice digits.}
    \label{fig:second}
\end{figure}

\section*{Results}

\subsection*{Optical experiment and agentic harness}

Our optical neural network realizes several diffractive layers with a single phase-only SLM and a mirror, between which the light is reflected several times before it reaches a camera~\cite{yildirim2024} (Fig.~\ref{fig:first}c). At each reflection the light meets a different region of the SLM, which acts as one layer of the network. The first region phase-encodes the input $x$, and the next two regions display the trainable phase masks $\varphi_1$ and $\varphi_2$, with several hundred thousand pixels in total. The camera records the output intensity, and for the classification of the handwritten digits 0, 1 and 2, the predicted class is the one of three detector regions on the camera that receives the most light. Although light propagates linearly between the planes, the output depends strongly nonlinearly on the parameters, because the input and the phase masks act on the phase of the field, the camera detects intensity and the SLM responds nonlinearly to its drive values, so that training is a non-convex problem. All methods start from phase masks trained by backpropagation through the ideal angular-spectrum model of this multi-reflection system, which classify 98\% of the test digits in simulation but only 35\% on the experiment, because the ideal model misses the imperfections of the device (Fig.~\ref{fig:first}a).

The agent works in an isolated workspace and reaches the physical optics only through the harness, which we designed to give it the working conditions of an experimentalist at the device (Fig.~\ref{fig:first}d, Methods). The probe tool displays any phase mask on any region of the SLM and returns the camera image, so that the agent can study the optics instead of inferring them from the task score. The evaluation and validation tools measure candidate phase masks on practice digits. Each request must state a hypothesis and an expected outcome, which makes every measurement a test of the agent's understanding of the optics. In addition, every camera image is charged to a common budget of measurements. The agent records its findings in a notes file, which serves as its memory and can be passed on to a later session. Its programs run on a central processing unit (CPU) only and the task forbids numerical models of light propagation, so that the agent has to learn about the device from its measurements. The gradient-free baselines run through the same harness, with the same start, budget and validations, and every method is scored once on sealed test digits.

The reasoning traces illustrate how the agent trains the network with these tools (Fig.~\ref{fig:first}e). The agent first verified the notes of the earlier session with a few probes. It then compared the camera images of the digit classes and noticed that the class of the input displaces the focused spots on the camera more than it changes their brightness. This effect follows from Fourier optics, because the local phase gradients of a phase-encoded digit tilt the wavefront and a lens converts each tilt into a shift of its focal spot, so that every class moves the spots in its own direction. To use this effect, the agent divided phase mask $\varphi_2$ into 49 blocks, measured the spot of every block and computed offline where each spot must land on the detectors. We refer to this design as the spot router, and its first version exceeded the final accuracy of the gradient-free methods after about a seventh of the budget.

\subsection*{Benchmark against gradient-free optimization}

On the optical classifier, the agent raised the accuracy to that of backpropagation through a calibrated model, whereas the gradient-free methods remained close to the start (Fig.~\ref{fig:second}). With the start phase masks, the camera image barely changes with the digit. With the design of the agent, each class sends the largest share of the detected light to its own detector and the confusion matrix is nearly diagonal (Fig.~\ref{fig:second}a,c--e). As baselines we used PPO, the best in situ training method in a recent comparison on optical processors~\cite{r21}, and its simpler variant, policy gradient. Their settings were first searched in simulation and the best deployed on the experiment (Methods), where they improved the test accuracy from 35\% to about 52\%. In contrast, our agent reached 91.1\% from the task description alone, and 96.8\% when it was given the notes and code of an earlier session, without any phase masks (Fig.~\ref{fig:second}b). This accuracy matches that of optical weights trained by backpropagation through a calibrated model of the device, 96.3\%. Such training is the most accurate that could be realized for this experiment, but for every device it requires a calibration campaign, here about 67,000 camera images, and many hours of training on graphics processors, whereas an agent run needs fewer measurements and no such training.

\begin{figure}[!t]
    \centerline{\includegraphics[width=18cm]{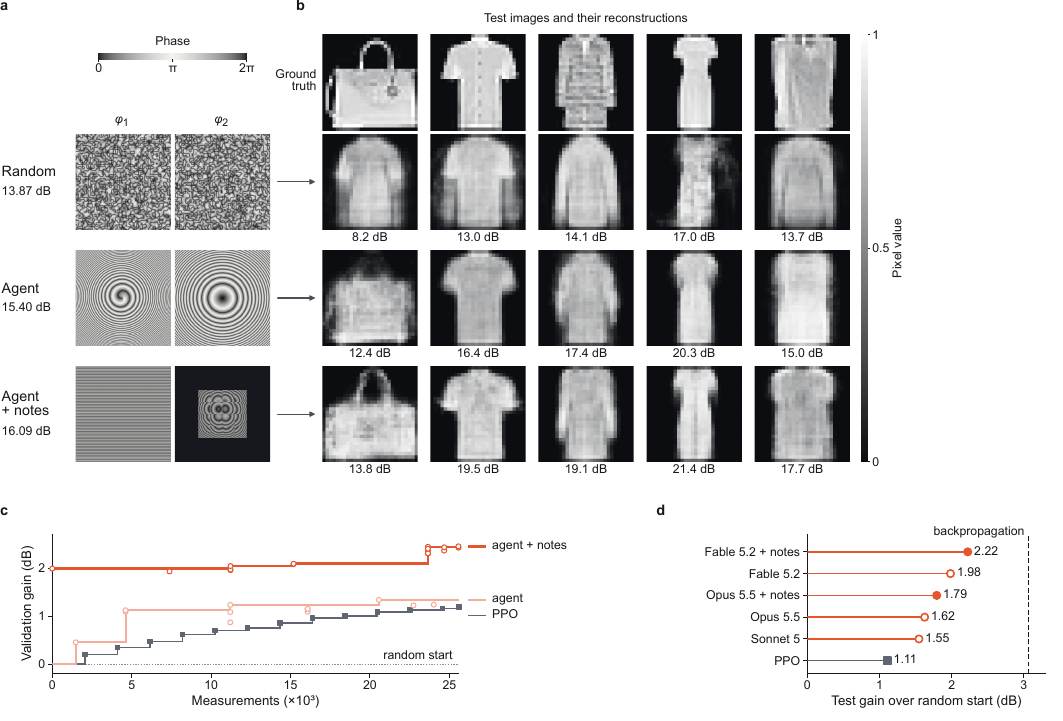}}
    \caption{\textbf{The agent trains an optical encoder that learns a compact representation of images.}
    The optics compress each Fashion-MNIST image into 16 detector signals, which play the role of the latent space of an autoencoder, and a digital decoder reconstructs the image from them.
    \textbf{a}, Phase masks $\varphi_1$ and $\varphi_2$ with random values, after an agent run without notes and after the agent run that received the notes and code of an earlier session, with the mean test peak signal-to-noise ratio (PSNR).
    \textbf{b}, Ground-truth test images (top) and their reconstructions by the digital decoder for the phase masks of \textbf{a} (arrows), with the PSNR of each reconstruction. The five images are spread evenly over the 50 test images that the agent run with notes improved most.
    \textbf{c}, Progress during training of the two agent runs of \textbf{a} and of PPO, the gradient-free baseline without a language model. Lines, gain in validation PSNR over the random start. Markers, single validations. The run with notes starts high because the agent rebuilt the earlier design before its first measurement.
    \textbf{d}, Gain in test PSNR over the random start at the end of the budget, one run per row, for three language models (Claude Fable~5.2, Opus~5.5 and Sonnet~5) and for PPO. Filled circles, runs with the notes and code of an earlier session. Open circles, runs with general lessons only. Square, PPO. Dashed, phase masks trained by backpropagation through a calibrated model.}
    \label{fig:third}
\end{figure}

Optically processed information usually has to be communicated, stored or processed further by electronics, and an optical front end that compresses an image into a few detector signals reduces the cost of this step~\cite{r38,r42,wang2023sensing}. We therefore tested whether the same harness can also train such an encoder (Fig.~\ref{fig:third}, Methods). In this task, the two phase masks compress each Fashion-MNIST image onto 16 detector regions of the camera, whose signals form a compact representation, like the latent space of an autoencoder, from which a small digital decoder reconstructs the image. The phase masks start from random values, and the quality of the reconstructions is measured by the peak signal-to-noise ratio (PSNR) on test images. As for the classifier, the agent first found a physical design, a lens on phase mask $\varphi_1$ that focuses the light into the grid of detectors, and then tuned a small set of aberration modes (Fig.~\ref{fig:third}a--c). With the same task text, Claude Fable 5.2, a more capable model~\cite{aaindex}, exceeded Claude Opus 5.5 and Claude Sonnet 5 (Fig.~\ref{fig:third}d), which indicates that the approach improves with the capability of the language model. Given also the notes and code of an earlier session, it reached 16.1~dB, against 15.0~dB for PPO and 13.8~dB for the random start (Fig.~\ref{fig:third}c,d). The best result obtained on this experiment, 16.9~dB, came from backpropagation through a calibrated model of the device, and the agent came within 0.8~dB of it without any calibrated model.

\subsection*{System identification and physical design}

\begin{figure}[p]
    \centerline{\includegraphics[width=18cm]{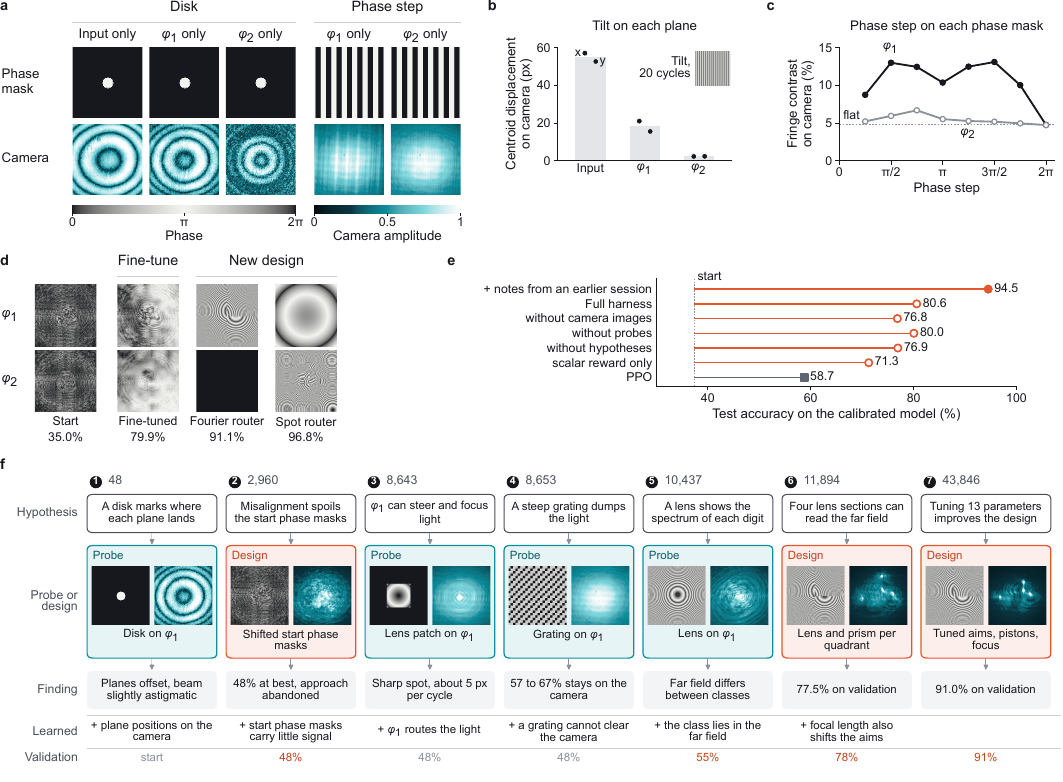}}
    \caption{\textbf{The agent understands the experiment through its own measurements and designs the optics from this understanding.}
    \textbf{a}, Probes chosen and written by the agent on the input and the two phase masks (top), and their camera images (bottom). Each probe occupies one plane, with the others flat. A disk locates each plane on the camera, and a phase step tests interference across each phase mask.
    \textbf{b}, Displacement of the intensity centroid on the camera for the same tilt (20 phase cycles) on each plane. Dots, $x$ and $y$ tilts.
    \textbf{c}, Fringe contrast for the same phase step on each phase mask. Dotted, flat.
    \textbf{d}, Both phase masks at the start and after the two routes of the runs, fine-tuning or a new design. The two new designs come from separate runs. The Fourier router (run from the task text alone) splits $\varphi_1$ into four lens sections. The spot router (run with earlier notes) focuses 49 blocks of $\varphi_2$ onto the detectors. Numbers, test accuracy.
    \textbf{e}, Test accuracy on the calibrated model with reduced information for the agent, and with earlier notes (filled circle). Square, PPO. Added rules did not help (Methods).
    \textbf{f}, Reasoning traces of the run from the task text alone, in seven steps from its first probe to the tuned Fourier router. Each step gives the hypothesis, the probe (blue) or design (red) that tested it, the finding, what the agent learned and the validation accuracy that followed. Top, measurements used (Methods).}
    \label{fig:fourth}
\end{figure}

Although the language model was trained neither on this experiment nor for optical system identification, the agents developed a physical understanding of the device by forming hypotheses and testing them with their own probes (Fig.~\ref{fig:fourth}a,f), and the quantities they measured agree with those of the calibrated model, with over a thousand times fewer measurements. This understanding consists of measured relationships, such as where each plane appears on the camera and how strongly each phase mask redirects the light, which the agents recorded in their notes (Fig.~\ref{fig:fourth}f). For example, the agents displayed the same tilt, a linear phase ramp, on each plane in turn and measured how far the light moved on the camera. The light moved most for a tilt on the input plane and least for one on phase mask $\varphi_2$ (Fig.~\ref{fig:fourth}b), and a phase step produced clear interference fringes on phase mask $\varphi_1$ but almost none on $\varphi_2$ (Fig.~\ref{fig:fourth}c). These measurements showed that light leaving phase mask $\varphi_2$ spreads only little over the short distance to the camera, whereas $\varphi_1$ and the input plane redistribute the light across the whole camera. The agents also found that the planes are offset from their nominal positions by several pixels and that the beam carries a slight astigmatism. These small deviations are the likely reason why the start phase masks fail on the experiment.

From such findings, each run chose one of two routes to a trained network (Fig.~\ref{fig:fourth}d). Some runs fine-tuned the start phase masks with a few physical parameters, such as plane shifts, lenses and low-order aberrations, and others designed new optics around the measured geometry. The reasoning traces of one of the latter show this from a blank page (Fig.~\ref{fig:fourth}f). The agent located the planes, abandoned fine-tuning and found that phase mask $\varphi_1$ steers the light. A lens on $\varphi_1$ then revealed a far field that differs between the digit classes, so the agent split $\varphi_1$ into four sections, each with its own lens and prism. It tuned this design, which we refer to as the Fourier router, without a model of the optics or an example design. In both routes, a small number of interpretable parameters replace several hundred thousand pixel updates, and each measurement therefore constrains the phase masks much more strongly than the measurement of a random perturbation.

\subsection*{Design of the harness}

Since the harness determines what the agent can learn about the device, we tested the contribution of each of its elements. For fast prototyping, we used the calibrated model of the experiment~\cite{r35}, on which many runs can proceed in parallel (Methods). On this model, we changed one element of the harness at a time (Fig.~\ref{fig:fourth}e). All elements that improved the accuracy give the agent more information about the optics, namely the camera images returned with every evaluation, probes at the cost of a single measurement, the hypothesis stated with every request and the notes file. In contrast, rules that prescribe how the agent should decide or spend its budget did not help, which suggests that the agent benefits more from better observations than from additional instructions. Even with the single scalar reward per candidate that the gradient-free methods receive, the agent clearly outperformed PPO. In this setting the agent is itself a gradient-free optimizer, but it interprets each result with what it knows about optics and devices, whereas PPO only tracks the statistics of the rewards. The information from the harness increases this advantage further. The largest gain came from notes passed on from an earlier session, without any phase masks, and such notes also produced the best runs on the experiment.

\section*{Discussion}

We have shown that a coding agent, working through a suitably designed harness, can train an optical neural network on the hardware from measurements alone. The agent took over the work that in situ training otherwise demands of an expert, from characterizing the device to designing and tuning its phase masks, and performed similarly to backpropagation through a calibrated model while requiring almost no expert involvement and no training on graphics processors. This result depends both on the harness, which defines what the agent can request and receive, and on the language model, a commercially available tool that contributes the physical reasoning and the choice of method. The agent thus brings an understanding of the optics, and as software it is consistent, always available and inexpensive.

These results point to an ability of language models that was not an explicit goal of their training. Without prior exposure to this device, the agents characterized its planes, misalignments and aberrations and designed a classifier and an encoder from these findings. In each case, they also provided a diagnosis that a physicist can check. Until now, language models have mostly assisted with programming and other digital knowledge work. Our results suggest that their understanding of optics is also sufficient to act on a real experiment, whose parameters can only be measured.

This ability is directly relevant for the deployment of optical neural networks. Since every device has its own imperfections, deploying thousands or millions of processors is impractical if each one needs an individually calibrated model or the attention of an expert. An agent could instead probe and train many devices in parallel, and return for each one its trained parameters together with an explanation of what it found. The notes serve this purpose, since they record in plain language what an agent learned about one device, which a later session can verify with a few probes. In our experiments, notes from a different session gave the largest gains, which shows that this knowledge transfers efficiently and could be shared between instances of the same or similar devices. The system demonstrated here can be regarded as a single layer of a larger network, and since the training does not need gradients through the device, many such units could be trained in the same way and combined. Beyond optical neural networks, the same approach could assist experimentalists with the alignment, calibration and tuning of other physical experiments. It is particularly promising for photonic systems that provide nonlinear and high-dimensional transformations, for example nonlinear propagation in multimode fibres~\cite{oguz2024prog}, lasers with optical feedback~\cite{r47} and light scattering in complex media~\cite{r48}, among other physical systems~\cite{r12}. Training neural networks with such systems has remained very challenging, because they are too complex to model accurately and their gradients are not available. Agents that learn about these systems from their own measurements could make them usable for efficient nonlinear computing, which may lead to new types of photonic processors.

\section*{Methods}
\paragraph{Experimental system.}

The multi-reflection optical system consists of a phase-only spatial light modulator (SLM, Meadowlark Optics HSP1920-500-1200, 1,920 \ensuremath{\times} 1,152 pixels) and a mirror, illuminated by a continuous-wave helium--neon laser at a wavelength of 632.8~nm. Light reflects from four 434 \ensuremath{\times} 434-pixel regions of the SLM. The first region is held flat and is not drawn in Fig.~\ref{fig:first}c, the second displays the input and the last two display the phase masks $\varphi_1$ and $\varphi_2$. A CMOS camera (IDS Imaging Development Systems UI-3240CP-M-GL) records a 384 \ensuremath{\times} 384-pixel amplitude image. Each measurement averages four camera frames after a settling time of 20~ms, at 0.107~s per measurement. For fast prototyping, we also replicated the experimental system in the digital domain, with a calibrated model based on the angular-spectrum method and fitted to measurements of the experiment~\cite{r35}. Phase masks trained by backpropagation through this calibrated model and scored on the experiment give the backpropagation reference. We also used the model to develop the harness and to select the settings of the baselines, and across the designs of the agent runs that we scored on both, the test accuracies on the experiment and on the model have a Pearson correlation of 0.88.

\paragraph{Classification task.}

Digits 0, 1 and 2 of the Modified National Institute of Standards and Technology (MNIST) data set are upsampled from 28 \ensuremath{\times} 28 to 200 \ensuremath{\times} 200 pixels and displayed with a phase of \(\pi\) times the pixel value. The predicted class is the detector, among three 48 \ensuremath{\times} 48-pixel regions of the camera, that receives the most light. The 2,000 practice digits come from the training split, and the 3,147 test digits are all test-split images of the three classes. The start phase masks were trained by backpropagation through the ideal model of the multi-reflection system, which uses angular-spectrum propagation and the camera registration of the experiment.

\paragraph{Agent and harness.}

Each run starts Claude Code, a program that lets a language model read and write files and run programs on a computer, with the language model Claude Opus 5.5. It runs without a person watching or typing (headless mode), in an isolated workspace. The reported runs used high reasoning effort, a predefined level of how long the model thinks between actions. For the comparison of language models on the encoder, the same harness also ran Claude Sonnet 5 and Claude Fable 5.2. These models span the current range of capability, from Claude Sonnet 5 at the lower end to Claude Fable 5.2, the highest-ranked model on the Artificial Analysis Intelligence Index at the time of writing~\cite{aaindex}, so that other current language models would be expected to perform within the range reported here. The programs of the agent run on a central processing unit (CPU) only. Access to graphics processors, to the network and to files outside the workspace is blocked, and the task text forbids numerical models of light propagation, so that the agent cannot simulate the optics and learns about them only from measurements. The budget is 51,200 measurements for the classifier and 25,600 for the encoder, where one measurement is one camera image of one displayed set of phase masks. In runs with notes from earlier sessions, the task text contained either the notes and code of one earlier run or a list of general lessons distilled from several runs, and never any phase masks or test data. To develop the harness on the calibrated model, we switched single elements on and off. Figure~\ref{fig:fourth}e shows the cases that change the information the agent receives. In these, we removed the camera images, so that the agent received only the detector signals, removed the probe tool, removed the requirement to state a hypothesis, or returned only one number per candidate, its reward, as the gradient-free methods receive. Additional rules on how to decide, how to spend the budget or how to review results did not raise the accuracy. The reasoning traces in Fig.~\ref{fig:fourth}f, the text and reasoning that the language model generated between its requests, were summarized from the full trace of the run from the task text alone with Claude Opus 5.5.

\paragraph{Baselines.}

Proximal policy optimization (PPO) samples full-resolution phase masks from a Gaussian policy around the current parameters, uses the clipped surrogate objective with four optimization epochs per batch, and learns the exploration width. Policy gradient uses antithetic sampling with normalized rewards. The reward is the negative cross-entropy of the temperature-scaled detector signals. To give the baselines their best conditions, we first searched their settings on the calibrated model of the experiment, in 22 settings that varied the exploration width, the learning rate, the batch shape and the number of PPO epochs. The best PPO setting was then used on the experiment, and policy gradient used a setting selected in the same search. Each baseline is validated every 2,560 measurements and keeps its best validated parameters, which matches the 20 validations of the agent.

\paragraph{Encoder task.}

Images of the Fashion-MNIST data set are displayed on the input region, and the two phase masks map them onto 16 detector regions of the camera, whose signals are normalized by the total intensity. A digital multilayer-perceptron decoder, trained beforehand together with a linear digital encoder of 16 outputs, reconstructs each image from the 16 measured signals through an affine map. The peak signal-to-noise ratio (PSNR) is computed on 300 held-out test images. Each method had 25,600 measurements and at most 20 validations. PPO, the stronger of the two baselines in the search of settings, used the classifier setting. The curves of Fig.~\ref{fig:third}c show the validations that each method requested during its run, on a fixed set of practice images, as the gain over the validation of the start phase masks in the same run. When the same random phase masks were measured again in the three sessions, their PSNR stayed about the same (13.84~dB, 13.82~dB and 13.87~dB), and each run is compared with the random phase masks of its own session.

\paragraph{Scoring and statistics.}

On the experiment, all parameter sets of one session were scored together in one interleaved block on the 3,147 test digits, so that even if the optics drift during the scoring, every method is affected equally. Because the test set is finite, each test accuracy carries a statistical uncertainty. For 3,147 digits, its standard error is at most 0.9 percentage points, the value for an accuracy of 50\%, and it is 0.3 points for the best run. The three classifier runs that received notes or general lessons from earlier sessions reached 89.2\%, 91.3\% and 96.8\% on the test digits (mean 92.4\%, standard deviation 3.9 percentage points). The duration of a run was set mainly by the acquisition rate of the camera, at 0.107~s per measurement or about 1.5~h for the classifier budget, and not by the language model. Since both the measurement budget and the reasoning effort are settings of the harness, the effort can be divided between measurements and reasoning to suit each setup. Where measurements are slow or costly, the agent can be asked to reason more and measure less, and where they are fast, to measure more. This flexibility could be useful for deploying inexpensive optical processors in large numbers, where the time and cost of training each device must stay small.

\bibliographystyle{naturemag}
\bibliography{refs}

\end{document}